\documentclass[11pt,a4paper]{article}
\usepackage{amsmath, amsfonts, amssymb}
\usepackage{a4wide}
\usepackage{parskip}
\usepackage{enumitem}
\usepackage{xcolor}

\usepackage{hyperref}
\usepackage{booktabs}
\usepackage{caption}
\usepackage{siunitx}
\usepackage{tabularx}
\usepackage{graphicx}
\usepackage{adjustbox}
\usepackage{multirow}
\usepackage{amsthm}
\def\proof{\noindent{\textit{Proof. }}}
\def\qed{\hfill {$\square$}\goodbreak \medskip}

\newtheorem{theorem}{Theorem}[section]
\newtheorem{lemma}[theorem]{Lemma}
\theoremstyle{definition}
\newtheorem{definition}[theorem]{Definition}
\newtheorem{example}[theorem]{Example}
\theoremstyle{conjecture}

\theoremstyle{proposition}
\newtheorem{proposition}[theorem]{Proposition}
\theoremstyle{remark}
\newtheorem{remark}[theorem]{Remark}

\theoremstyle{corollary}

\numberwithin{equation}{section}

\usepackage{lipsum} 

\usepackage{tikz,xcolor,hyperref}

\definecolor{lime}{HTML}{A6CE39}
\DeclareRobustCommand{\orcidicon}{%
	\begin{tikzpicture}
		\draw[lime, fill=lime] (0,0) 
		circle [radius=0.16] 
		node[white] {{\fontfamily{qag}\selectfont \tiny ID}};
		\draw[white, fill=white] (-0.0625,0.095) 
		circle [radius=0.007];
	\end{tikzpicture}
	\hspace{-2mm}
}

\foreach \x in {A, ..., Z}{%
	\expandafter\xdef\csname orcid\x\endcsname{\noexpand\href{https://orcid.org/\csname orcidauthor\x\endcsname}{\noexpand\orcidicon}}
}

\begin{document}
	\date{}
	{\vspace{0.01in}
		\title{Codes over the non-unital non-commutative ring $E$ using simplicial complexes}
		\author{{\bf Vidya Sagar\footnote{email: {\tt vsagariitd@gmail.com}}\orcidA{}  \;and  \bf Ritumoni Sarma\footnote{	email: {\tt ritumoni407@gmail.com}}\orcidB{}} \\ Department of Mathematics,\\ Indian Institute of Technology Delhi,\\Hauz Khas, New Delhi-110016, India. }
		\maketitle
		\begin{abstract}			
			There are exactly two non-commutative rings of size $4$, namely, $E = \langle a, b ~\vert ~ 2a = 2b = 0, a^2 = a, b^2 = b, ab= a, ba = b\rangle$ and its opposite ring $F$. These rings are non-unital. A subset $D$ of $E^m$ is defined with the help of simplicial complexes, and utilized to construct linear left-$E$-codes $C^L_D=\{(v\cdot d)_{d\in D} : v\in E^m\}$ and right-$E$-codes $C^R_D=\{(d\cdot v)_{d\in D} : v\in E^m\}$. We study their corresponding binary codes obtained via a Gray map. The weight distributions of all these codes are computed. We achieve a couple of infinite families of optimal codes with respect to the Griesmer bound. Ashikhmin-Barg’s condition for minimality of a linear code is satisfied by most of the binary codes we constructed here. All the binary codes in this article are few-weight codes, and self-orthogonal codes under certain mild conditions. This is the first attempt to study the structure of linear codes over non-unital non-commutative rings using simplicial complexes.
			
			\medskip
			
			
			\medskip
			
			
		\end{abstract}
		\section{INTRODUCTION}
		The study of codes over the rings of size four, namely, $\mathbb{Z}_4$ (in \cite{Z4}), $\mathbb{F}_4$ (in \cite{F4}), $\mathbb{F}_2\times \mathbb{F}_2$ (in \cite{F2xF2}), $\mathbb{F}_2 + u\mathbb{F}_2$ (in \cite{F2uF2}), have got huge attention in the area of algebraic coding theory in the past. These are the only  commutative unital rings among the 11 rings of size four classified by Fine \cite{Fine}. Among these 11 rings, 5 of them are commutative non-unital (namely, $B$, $C$, $H$, $I$, $J$) and 2 of them are non-commutative non-unital (namely, $E$, $F$). In 2020, Alahmadi et. al. \cite{AlahmadiE1} considered the non-commutative non-unital ring $E$ (see \cite{Fine}) in the context of QSD codes for the first time. They studied a multilevel construction of a quasi self dual (QSD) code as a function of a pair of dual codes, and classified QSD codes of length $n \leq 6$. Following the work in \cite{AlahmadiE1}, the authors in \cite{AlahmadiE2} studied construction of self-orthogonal codes over $E$ and classified QSD codes of length $n \leq 12$. Since then codes over non-unital rings have been studied by many researchers (see \cite{AlahmadiI2, AlahmadiI1, DNAnonunital, KimI, MinjaNonunital, LCDnonUnital}).\par		
		In this article, we study the linear left-$E$-code $C^L_D = \{(v\cdot d)_{d\in D} : v\in E^m\}$ and right-$E$-code $C^R_D = \{(d\cdot v)_{d\in D} : v\in E^m\}$, where $D\subseteq E^m$ and $m$ is a positive integer. The authors in \cite{Ding} first introduced the above construction of $C_D$ in order to generalize the Mattson-Solomon transform for cyclic codes. If the defining set $D$ is constructed using simplicial complexes, the study of $C_D$ become convenient. In fact, several interesting linear codes have been constructed in the recent past by using simplicial complexes (see \cite{Hyun_Kim, Hyun_Lee, Sagar_Sarma2, Sagar_Sarma3,shi_x, shi_nonchain,  shi_x2, Shi_guan, shi_qian, shi_xuan, mixed2, wu_zhu, Zhu_Wei}). It is expected that for properly chosen finite fields (more generally, rings) and defining sets, we may be able to discover codes with good parameters.\par 
		Yansheng et. al. in \cite{Wu_Li} studied linear codes over $\mathbb{F}_{4}$ and their binary subfield codes, and obtained the weight distributions of all these codes. They produced two infinite families of optimal linear codes. Motivated by this work, the authors in \cite{Sagar_Sarma} studied octanary linear codes using simplicial complexes, and obtained minimal and optimal codes. Recently, the authors in \cite{GeneralCase} generalized the work of \cite{Sagar_Sarma} for finite fields of characteristic $2$ by using LFSR sequences. The weight distribution (see \cite{wchuffman}) of a linear code contains crucial information regarding error detecting as well as error correcting capacity of the code, and it allows the computation of the error probability of error detection and correction with respect to some algorithms \cite{Klove}. Few-weight codes are useful because of their connection with strongly regular graphs, difference sets and finite geometry \cite{Few1, Few2}. The minimum Hamming distance of linear codes are well known for their importance in determining error-correcting capacity. As a result, finding optimal linear codes has become one of the central topics for researchers. In \cite{Hyun_Lee}, the authors showed how optimal codes can be utilized for the construction of secret sharing schemes with nice access structures following the framework discussed in \cite{YuanDing}. Minimal codes are useful to construct the access structure of the secret sharing schemes \cite{Ding_Ding, MasseyMinimal}. This special class of codes is also important as these can be decoded by using the minimum distance decoding rule \cite{suffConMinimalcode}. Normally, it is difficult to identify all the minimal codewords of a given code even over a finite field of characteristic $2$. In view of this, researchers began to investigate minimal codes.\par 		
		Inspired by the work of \cite{Wu_Li}, a natural endeavour is to explore the construction of linear codes over non-unital rings with the help of simplicial complexes and investigate the structure of these codes. We consider the rings $E$ and $F$ according to the notation of Fine \cite{Fine}. We, in this article, choose a defining set $D$ with the help of certain simplicial complexes, and study the structure of linear left-$E$-codes and right-$E$-codes. We obtain the Lee-weight distributions for the codes over $E$ by using Boolean functions. By considering a Gray map, we obtain certain binary linear codes, and their weight distributions. These binary codes turn out to be self-orthogonal under certain mild conditions. By choosing the defining set appropriately, we produce two infinite families of optimal codes. Moreover, most of the binary codes obtained here are minimal. We give a few examples that support our results.\par
		The remaining sections of this article are arranged as follows. Preliminaries are presented in the next section. By using simplicial complexes, linear left-$E$-codes and their binary Gray images are studied in Section \ref{section3}. Linear right-$E$-codes and their binary Gray images are investigated in Section \ref{section5}. Section \ref{section6} concludes this article.

		\section{Definitions and Preliminaries}\label{section2}
		Let $E$ and $F$ (in the notation of Fine \cite{Fine}) be the rings given by $E = \langle a, b ~\vert ~ 2a = 2b = 0, a^2 = a, b^2 = b, ab= a, ba = b\rangle $ and $F = \langle a, b ~ \vert ~ 2a = 2b = 0, a^2 = a, b^2 = b, ab = b, ba = a \rangle $. Note that the underlying sets of $E$ and $F$ are both $\{0, a, b, c = a+b\}$. Both $E$ and $F$ are non-commutative and non-unital; and $F$ is the opposite ring of $E$. The addition and multiplication tables of $E$ are given as follows:
		\begin{center}			
			\begin{tabular}{|c|c|c|c|c|}
				\hline
				$+$ & 0 & a & b & c \\
				\hline
				0 & 0 & a & b & c \\
				\hline
				a & a & 0 & c & b \\
				\hline
				b & b & c & 0 & a \\
				\hline
				c & c & b & a & 0 \\
				\hline
			\end{tabular}
			~~~~~~~~~~~~~~~~~~~~~~~~~
			\begin{tabular}{|c|c|c|c|c|}
				\hline
				$\times$ & 0 & a & b & c \\
				\hline
				0 & 0 & 0 & 0 & 0 \\
				\hline
				a & 0 & a & a & 0 \\
				\hline
				b & 0 & b & b & 0 \\
				\hline
				c & 0 & c & c & 0 \\
				\hline
			\end{tabular}~~~
		\end{center}
		Consider the following action of $\mathbb{F}_2$ on $E$: $e0 = 0e = 0$ and $e1 = 1e = e$ for all $e \in E$. Then every element of $E$ can be expressed as $as+ct$ for $s, t \in \mathbb{F}_2$.
		\begin{lemma}
			For $n \in \mathbb{N}$, $E^n = a\mathbb{F}_2^n + c\mathbb{F}_2^n$ and the sum is direct.	    	
		\end{lemma}		        
        Suppose $\mathcal{R} = E$ (or $F$).
		Let $\Phi : \mathcal{R} \longrightarrow \mathbb{F}_2^2$ be the \textit{Gray map} given by 
		\begin{equation}\label{PhiEq}
			\Phi(as + ct) = (t, s+t).
		\end{equation}
		This extends to a map from $\mathcal{R}^m$ to $(\mathbb{F}_2^{2})^m$ component-wise for any $m \in \mathbb{N}$.
		\begin{definition}
			A \textit{linear left-$E$-code} (respectively, \textit{right-$E$-code}) of length $m$ is a left-$E$-submodule (respectively, right-$E$-submodule) of $E^m$.
		\end{definition}
	\begin{remark}
		Every right-$E$-module is a left-$F$-module and conversely. Similarly, a left-$E$-module and a right-$F$-module are same. Therefore, we shall study linear left-$E$-codes and right-$E$-codes only, and will not write any assertions for codes over the ring $F$.
	\end{remark}
		Let $v, w \in \mathbb{F}_2^m$. Then the \textit{Hamming weight} of $v$ denoted by $wt_{H}( v )$ is the number of non-zero entries in $v$. The \textit{Hamming distance} between $v$ and $w$ is $d_H(v, w) = wt_{H}(v-w)$.\\
		Let $x=a\alpha +c\beta$, $y=a\alpha' + c\beta' \in E^m$, where $\alpha, \beta, \alpha', \beta'\in \mathbb{F}_2^m$. Then the \textit{Lee weight} of $x$ is $wt_{Lee}(x) = wt_{H}(\Phi(x)) = wt_{H}(\beta) + wt_{H}(\alpha +\beta )$. The \textit{Lee distance} between $x$ and $y$ is $d_{L}(x, y) = wt_{Lee}(x-y)$. Thus, the map $\Phi$ is an isometry. Note that the image of a linear left-$E$-code (respectively, right-$E$-code) under the above Gray map is a binary linear code. Suppose $C$ is a linear $E$-code of length $m$. Let $A_i$ be the cardinality of the set that contains all codewords of $C$ having Lee weight $i$, $0\leq i\leq 2m$. Then the homogeneous polynomial in two variables \[Lee_C(X, Y) = \sum_{c\in C}X^{2m-wt_{Lee}(c)}Y^{wt_{Lee}(c)}\] is called the \textit{Lee weight enumerator} of $C$ and the string $(1, A_1,\dots ,A_{2m})$ is called the \textit{Lee weight distribution} of $C$. In a similar way, we can define Hamming weight enumerator and Hamming weight distribution of a linear code over a finite field. In addition, if the total number of $i\geq 1$ such that $A_i\neq 0$ is $l$, then $C$ is called an $l$-\textit{weight linear code}. Every $1$-weight linear code is an equidistant linear code. Bonisoli characterized all equidistant linear codes over finite fields in \cite{Bonisoli}.
       \begin{theorem}\cite{Bonisoli}\textnormal{(Bonisoli)}\label{Bonisoli}
       	Suppose $C$ is a equidistant linear code over $\mathbb{F}_q$. Then $C$ is equivalent to the $r$-fold replication of a simplex code, possibly with added $0$-coordinates.
       \end{theorem}
       An $[n, k, d]$-linear code $C$ is called \textit{distance optimal} if there exist no $[n,k,d+1]$-linear code (see \cite{wchuffman}). Next we recall the Griesmer bound.
       \begin{lemma}\label{griesmerbound}\cite{Griesmer}
       	\textnormal{(Griesmer Bound)} If $C$ is an $[n,k,d]$-linear code over $\mathbb{F}_q$, then we have
       	\begin{equation}\label{GriesmerBound}
       		\sum\limits_{i=0}^{k-1}\left\lceil \frac{d}{q^i}\right\rceil \leq n,
       	\end{equation}  	
       	where $\lceil \cdot \rceil$ denotes the ceiling function.
       \end{lemma}
       A linear code is called a \textit{Griesmer code} if equality holds in Equation \eqref{GriesmerBound}. Note that every Griesmer code is distance optimal, but the converse need not be true.\\
       For $m\in \mathbb{N}$, we shall write $[m]$ to denote the set $\{1, 2,\dots ,m\}$ and $w\in \mathbb{F}_{2}^m$. Then the set $\textnormal{Supp}(w)=\{ i\in [m]: w_i = 1\}$ is called the \textit{support} of $w$. Note that the Hamming weight of $w\in \mathbb{F}_2^m$ is $wt_H(w)=\vert \textnormal{Supp}(w)\vert $. For $v,w\in \mathbb{F}_{2}^m$, one says that $v$ \textit{covers} $w$ if $\textnormal{Supp}(w)\subseteq \textnormal{Supp}(v)$. If $v$ covers $w$, we write $w\preceq v$.\\
	Consider the map $\psi: \mathbb{F}_2^m\longrightarrow 2^{[m]}$ is defined as $\psi(w)=\textnormal{Supp}(w)$, where $2^{[m]}$ denotes the power set of $[m]$. Note that $\psi$ is a bijective map. Now onwards, we will write $w$ instead of Supp($w$) whenever we require.
	\begin{definition}
		Let $C$ be a linear code over $\mathbb{F}_2$. An element $v\in C\setminus \{0\}$ is called \textit{minimal} if $w\preceq v$ and $w\in C\setminus \{0\}$ $\implies$ $w = v$. If each nonzero codeword of $C$ is minimal then $C$ is called a \textit{minimal code}.
	\end{definition}
	Now we recall a result from \cite{suffConMinimalcode} on the minimality of a code over a finite field.
	\begin{lemma}\label{minimal_lemma}
		\cite{suffConMinimalcode}\textnormal{(Ashikhmin-Barg)}
		Let $C$ be a linear code over $\mathbb{F}_q$ with $wt_o$ and $wt_\infty$ as minimum and maximum weights of its non-zero codewords. If $\frac{wt_0}{wt_{\infty}}> \frac{q-1}{q}$, then $C$ is minimal.
	\end{lemma}
	\begin{definition}
		A subset $\Delta$ of $\mathbb{F}_{2}^m$ is called a \textit{simplicial complex} if $v\in \Delta, w\in \mathbb{F}_2^m$ and $w\preceq  v$ $\implies$ $w\in \Delta$. An element $v\in \Delta$ is called a \textit{maximal element of} $\Delta$ if for $w\in \Delta$, $v\preceq w$ $\implies$ $v=w$.
	\end{definition} 
    A simplicial complex can have more than one maximal elements. Let $M\subseteq [m]$. The simplicial complex generated by $M\subseteq [m]$ is denoted by $\Delta_{M}$ and is defined as 
	\begin{equation}
		\Delta_{M}=\{w\in \mathbb{F}_{2}^m \vert \textnormal{ Supp}(w)\subseteq M\}=\psi^{-1}(2^M).
	\end{equation}
	Note that $\psi^{-1}(M)$ is the only maximal element of $\Delta_M$, and $\vert \Delta_{M}\vert =\vert 2^M\vert=2^{\vert M\vert}$. Here $\Delta_{M}$ is a vector space over $\mathbb{F}_2$ of dimension $\vert M \vert$.\par
	Given a subset $P$ of $\mathbb{F}_{2}^m$, define the polynomial (referred to as an $m$-variable generating function, see \cite{Chang}) $\mathcal{H}_{P}(y_1, y_2,\dots , y_m)$ by
	\begin{equation}
		\mathcal{H}_P(y_1,y_2,\dots ,y_m)=\sum\limits_{v\in P}\prod_{i=1}^{m}y_i^{v_i}\in \mathbb{Z}[y_1,y_2,\dots ,y_m],
	\end{equation}
	where $v=(v_1,\dots ,v_m)$ and $\mathbb{Z}$ denotes the ring of integers.\\	
	We recall a lemma from \cite{Chang}.
	\begin{lemma}\cite{Chang}\label{generatinglemma}
		Suppose $\Delta\subseteq \mathbb{F}_{2}^m$ is a simplicial complex and $\mathcal{F}$ consists of its maximal elements. Then 
		\begin{equation}
			\mathcal{H}_{\Delta}(y_1,y_2,\dots ,y_m)=\sum\limits_{\emptyset\neq S\subseteq\mathcal{F}}(-1)^{\vert S\vert +1}\prod_{i\in \cap S}(1+y_i),
		\end{equation}
		where $\cap S=\bigcap\limits_{F\in S}\textnormal{Supp}(F)$. In particular, we have
		\begin{equation*}
			\vert \Delta\vert =\sum\limits_{\emptyset\neq S\subseteq \mathcal{F}}(-1)^{\vert S\vert +1}2^{\vert \cap S\vert}.
		\end{equation*}
	\end{lemma}
	\begin{example}
		Consider the simplicial complex
		\begin{equation*}
			\Delta = \{(0, 0, 0, 0), (1, 0, 0, 0), (0, 0, 1, 0), (1, 0, 1, 0), (0, 1, 0, 0), (0, 0, 0, 1), (0, 1, 0, 1)\}.
		\end{equation*}
		Then $\mathcal{F} = \{F_1, F_2\}$ where $F_1 = (1, 0, 1, 0)$ and $F_2 = (0, 1, 0, 1)$. So 
		\begin{equation*}
			\begin{split}
				\mathcal{H}_{\Delta}(y_1, y_2, y_3, y_4) & = \prod_{i\in \textnormal{Supp}(F_1)}(1+y_i) + \prod_{i\in \textnormal{Supp}(F_2)}(1+y_i) -1\\
				& = 1 + y_1 + y_3 + y_1y_3 + y_2 + y_4 + y_2y_4.				
			\end{split}
		\end{equation*}
	 and $\vert \Delta \vert = 7.$
	\end{example}
    For $M\subseteq [m]$, define a Boolean function $\Psi(\cdot \vert M): \mathbb{F}_2^m \longrightarrow \mathbb{F}_2$  by
    \begin{equation}\label{BooleanFunction}
    	\Psi(\alpha\vert M)=\prod_{i\in M}(1-\alpha_i)=\begin{cases}
    		1, \text{  if } \textnormal{ Supp}(\alpha)\cap M=\emptyset;\\
    		0, \text{  if } \textnormal{ Supp}(\alpha)\cap M\neq \emptyset.
    	\end{cases}
    \end{equation}
    Here we recall a lemma from \cite{Sagar_Sarma2}.
    \begin{lemma}\cite{Sagar_Sarma2}\label{countingLemma}
    	Suppose $M, N$ are subsets of $[m]$. Then
    	\begin{enumerate}
    		\item \begin{enumerate}
    			\item \label{1.1}
    			$\vert \{v\in \mathbb{F}_{2}^m : \Psi(v \vert M) = 1\}\vert = 2^{m-\vert M\vert}$.
    			\item \label{1.2}
    			$\vert \{v\in \mathbb{F}_{2}^m : \Psi(v \vert M) = 0\}\vert = (2^{\vert M\vert }-1)\times 2^{m-\vert M\vert}$.
    		\end{enumerate}    		
    		\item \label{2.0}
    		$\vert \{v\in \mathbb{F}_2^m : \Psi(v\vert M )=0, \Psi(v\vert N) = 0\}\vert = (2^{\vert M\vert }-1)\times 2^{m-\vert M\vert } + (2^{\vert N\vert }-1)\times 2^{m-\vert N\vert }-(2^{\vert M\cup N\vert }-1)\times 2^{m-\vert M\cup N\vert }$.
    		
    		\item \label{3.0}
    		\begin{enumerate}
    			\item $\vert \{(v, w)\in \mathbb{F}_2^m \times \mathbb{F}_2^m : v \neq w, \Psi(w\vert M )=0, \Psi(v+w\vert M) = 0\}\vert = \{(2^{\vert M \vert} -2)\times 2^{m-\vert M \vert} + (2^{m-\vert M \vert} - 1)\}\times (2^{\vert M \vert } -1)\times 2^{m-\vert M \vert}$.
    		    
    		    \item $\vert \{(v, w)\in \mathbb{F}_2^m \times \mathbb{F}_2^m : v \neq w, \Psi(w\vert M )=0, \Psi(v+w\vert M) = 1\}\vert = \{(2^{\vert M \vert} -1)\times 2^{m-\vert M \vert}\}\times (2^{m - \vert M \vert } -1)$.
    		    
    		    \item $\vert \{(v, w)\in \mathbb{F}_2^m \times \mathbb{F}_2^m : v \neq w, \Psi(w\vert M )=1, \Psi(v+w\vert M) = 0\}\vert = \{(2^{\vert M \vert} -1)\times 2^{m-\vert M \vert}\}\times (2^{m - \vert M \vert } -1)$.
    		    
    		    \item $\vert \{(v, w)\in \mathbb{F}_2^m \times \mathbb{F}_2^m : v \neq w, \Psi(w\vert M )=1, \Psi(v+w\vert M) = 1\}\vert = \{ (2^{m-\vert M \vert} -1)\}\times (2^{m - \vert M \vert } -2)$.

    		\end{enumerate}

    	\end{enumerate}
    \end{lemma}    
    
    Here we recall a result from \cite{shi_nonchain}.
    \begin{lemma}\cite{shi_nonchain}\label{relation_with_complement}
    	Let $\alpha \in \mathbb{F}_2^m$ and let $\Delta_{M}$ be the simplicial complex generated by $M \subseteq [m]$. Then
    	\begin{equation*}
    		\begin{split}
    			\sum_{t\in \Delta_{M}^{\textnormal{c}}}(-1)^{\alpha t}=2^{m}\delta_{0, \alpha}-\sum_{t\in \Delta_{M}}(-1)^{\alpha t},
    		\end{split}
    	\end{equation*}
    	where $\delta$ is the Kronecker delta function, and $\Delta_{M}^{\textnormal{c}}=\mathbb{F}_2^m\setminus \Delta_{M}$, the complement of $\Delta_{M}$.
    \end{lemma}
	In the forthcoming sections, we study the algebraic structures of linear left-$E$-codes (respectively, right-$E$-codes) and their Gray images.
	\section{Linear left-$E$-codes using simplicial complexes}\label{section3}
	  For any $n, m\in \mathbb{N}$, let $D=\{d_1<d_2<\dots <d_n\}$ be an ordered multiset, where $d_i \in E^m$ $\forall$ $i$.
	  Let $m\in \mathbb{N}$ and let $D_i\subseteq \mathbb{F}_2^m, i=1, 2$. Assume that $D=aD_1+cD_2\subseteq E^m$. Define
		\begin{equation}\label{C_Ddefinitionleft}
			C^L_D=\{c^L_D(v)=\big(v\cdot d\big)_{d\in D} ~\vert~ v \in E^m\}.
		\end{equation}
	   where $x\cdot y=\sum\limits_{i=1}^{m}x_iy_i$ for $x,y\in E^m$.\\
	   Then $C^L_D$ is a linear left-$E$-code of length $\vert D\vert $. The ordered set $D$ is called the \textit{defining set} of $C^L_{D}$. Throughout this article we consider a defining set to an ordered multiset. Note that on changing the order of $D$ we will get a code which is permutation equivalent (see \cite{wchuffman}) to $C^L_D$.\\	   
	   Observe that $c^L_{D}: E^m\longrightarrow C^L_{D}$ defined by $c^L_{D}(v)=\big(v\cdot d\big)_{d\in D}$ is an epimorphism of left-$E$-modules.
	   \subsection{Weight distributions of linear left-$E$-codes}
	   
	   Assume that $x=a\alpha + c\beta \in E^m$ and $d=at_1 +ct_2\in D$, where $\alpha, \beta \in \mathbb{F}_2^m$ and $t_i\in D_i, i=1, 2$. Then the Lee weight of $c^L_{D}(x)$ is 
	   \begin{equation*}
	   	\begin{split}
	   		wt_{Lee}(c^L_{D}(x)) = & wt_{Lee}\big(\big(\big(a\alpha +c \beta \big)\cdot \big(at_1 + ct_2 \big)\big)_{t_i\in D_i}\big)\\
	   		                 = & wt_{Lee}\big(\big(a(\alpha t_1) + c(\beta t_1)\big)_{t_i\in D_i}\big)\\
	   		                 = & wt_{H}\big(\big(\beta t_1\big)_{t_i\in D_i}\big) + wt_{H}\big(\big(\alpha t_1 + \beta t_1\big)_{t_i\in D_i}\big)
	   	\end{split}
	   \end{equation*}   
       Now if $v \in \mathbb{F}_{2}^m$, then $wt_H(v)=0 \iff v=\textbf{0}\in \mathbb{F}_{2}^m $. Hence,
       \begin{equation}\label{keyeq1}
   	     \begin{split}
   		   wt_{Lee}(c^L_{D}(x))& =  \vert D \vert -\frac{1}{2}\sum\limits_{t_1\in D_1}\sum\limits_{t_2\in D_2}\big(1+(-1)^{\beta t_1 } \big)\\
   		    &+\vert D \vert -\frac{1}{2}\sum\limits_{t_1\in D_1}\sum\limits_{t_2\in D_2}\big(1+(-1)^{(\alpha +\beta ) t_1} \big)\\
   		    &=\vert D\vert -\frac{1}{2}\sum\limits_{t_1\in D_1}(-1)^{\beta t_1}\sum\limits_{t_2\in D_2}(1) - \frac{1}{2} \sum\limits_{t_1\in D_1}(-1)^{(\alpha + \beta )t_1}\sum\limits_{t_2\in D_2}(1).
   	    \end{split}
       \end{equation}
       For $Q \subseteq \mathbb{F}_2^m$ and $\alpha \in \mathbb{F}_2^m$, we define
       \begin{equation}\label{chi_def}
       	 \chi_{\alpha}(Q) = \sum\limits_{t\in Q}(-1)^{\alpha t}.
       \end{equation}
       Note that, for $\alpha \in \mathbb{F}_2^m$ and $\emptyset \neq M \subseteq [m]$, we have
       \begin{equation}\label{keyeq2}
       	   \begin{split}
       	   	  \chi_{\alpha}(\Delta_{M})&=\sum\limits_{t\in \Delta_{M}}(-1)^{\alpha t}\\
       	   	  &= \mathcal{H}_{\Delta_{M}}\big((-1)^{\alpha_1}, (-1)^{\alpha_2},\dots , (-1)^{\alpha_m}\big)\\
       	   	  &=\prod_{i\in M}(1+(-1)^{\alpha_i})=\prod_{i\in M}(2-2\alpha_i) \textnormal{ (By Lemma }\ref{generatinglemma})\\
       	   	  &=2^{\vert M \vert }\prod_{i\in M}(1-\alpha_i) = 2^{\vert M \vert}\Psi(\alpha\vert M),
       	   \end{split}
       \end{equation}
      where $\Psi(\cdot \vert M)$ is the Boolean function defined in Equation (\ref{BooleanFunction}).\\
      The following result describes Lee weight distributions of $C^L_D$ for various choices of $D$.
		\begin{theorem}\label{main}
			Suppose that $m\in \mathbb{N}$ and $M, N\subseteq [m]$.
			\begin{enumerate}
				\item \label{proposition:1}
				Let $D=a\Delta_M+ c\Delta_N\subseteq E^m$. Then $C^L_D$ is a linear left-$E$-code of length $\vert D\vert = 2^{\vert M\vert + \vert N \vert }$ and size $2^{2\vert M \vert}$. The Lee weight distribution of $C^L_D$ is displayed in Table \ref{table:1}.
					\begin{table}[]
						\centering
							\begin{tabular}{  c | c  }
								\hline
								Lee weight    & Frequency \\
								\hline
								$2^{\vert M \vert + \vert N \vert }$ & $2^{2m-2\vert M\vert}(2^{\vert M \vert}-1)^2$\\
								\hline
								  $2^{ \vert M \vert + \vert N \vert -1}$ & $2^{2m-2\vert M\vert +1}(2^{\vert M\vert }-1)$\\     				
								\hline
								$0$ & $2^{2m-2\vert M\vert}$\\
								\hline
							\end{tabular}
						\caption{Lee weight distribution in Theorem \ref{main} (\ref{proposition:1})}
						\label{table:1}	
					\end{table}

				\item \label{proposition:2}
				Let $D=a\Delta^{c}_M + c\Delta_N \subseteq E^m$. Then $C^L_D$ is a linear left-$E$-code of length $\vert D\vert = (2^m-2^{\vert M\vert})\times 2^{ \vert N \vert}$ and size $2^{2m}$. The Lee weight distribution of $C^L_D$ is displayed in Table \ref{table:2}.
					\begin{table}[]
						\centering
						\begin{adjustbox}{width=\textwidth}
							\begin{tabular}{  c | c  }
								\hline
								Lee weight    & Frequency \\
								\hline
								$ 2^{m+\vert N\vert}$ & $2^{2m-2\vert M \vert} - 2^{m-\vert M\vert +1} + 1$\\
								\hline
								$2^{m+ \vert N\vert - 1}$ & $2^{m-\vert M \vert+1}-2$\\
								\hline
								$2^{m+\vert N \vert }-2^{\vert M\vert + \vert N \vert }$ & $2^{2m}+2^{2m-2\vert M \vert }-2^{2m-\vert M\vert +1}$\\
								\hline
								$2^{m + \vert N\vert} -2^{\vert M\vert + \vert N \vert -1}$ & $2^{2m-\vert M\vert +1} - 2^{2m-2\vert M\vert +1}- 2^{m+1} + 2^{m- \vert M \vert +1}$\\
								\hline
								$2^{m+\vert N \vert -1} - 2^{\vert M \vert + \vert N \vert -1}$ & $2^{m + 1} - 2^{m-\vert M \vert +1}$\\
								\hline
								$0$ & $1$\\
								\hline
							\end{tabular}
						\end{adjustbox}
						\caption{Lee weight distribution in Theorem \ref{main} (\ref{proposition:2})}
						\label{table:2}		
					\end{table}
				
				\item \label{proposition:3}
				Let $D=a\Delta_M + c\Delta^{c}_N\subseteq E^m$. Then $C^L_D$ is a linear left-$E$-code of length $\vert D\vert = 2^{\vert M \vert}\times(2^m-2^{\vert N \vert})$ and size $2^{2\vert M\vert }$. The Lee weight distribution of $C^L_D$ is displayed in Table \ref{table:3}.
				\begin{table}[]
					\centering
						\begin{tabular}{  c | c  }
							\hline
							Lee weight    & Frequency \\
							\hline
							$ 2^{\vert M \vert}\times (2^{m}- 2^{\vert N\vert})$ & $2^{2m} - 2^{2m-\vert M\vert +1}+2^{2m-2\vert M\vert}$\\
							\hline
							$2^{\vert M \vert -1}\times (2^{m}-2^{\vert N \vert})$ & $2^{2m-\vert M\vert +1}-2^{2m-2\vert M\vert +1}$\\
							\hline
							$0$ & $ 2^{2m - 2\vert M \vert }$\\
							\hline
						\end{tabular}
					\caption{Lee weight distribution in Theorem \ref{main} (\ref{proposition:3})}
					\label{table:3}		
				\end{table}
				
				\item \label{proposition:4} 
				Let $D=a\Delta^{c}_M + c\Delta_N^{c}\subseteq E^m$. Then $C^L_D$ is a linear left-$E$-code of length $\vert D\vert = (2^m-2^{\vert M\vert })(2^m-2^{\vert N\vert })$ and size $2^{2m}$. The Lee weight distribution of $C^L_D$ is displayed in Table \ref{table:4}.
				\begin{table}[]
					\centering
					\begin{adjustbox}{width=\textwidth}
						\begin{tabular}{  c | c  }
							\hline
							Lee weight    & Frequency \\
							\hline
							$2^{m}\times (2^m-2^{\vert N\vert})$ & $2^{2m-2\vert M\vert }-2^{m-\vert M\vert +1} +1$\\
							\hline
							$2^{m-1}\times (2^m-2^{\vert N\vert})$ & $2^{m-\vert M\vert +1}-2$\\
							\hline
							$(2^m-2^{\vert M \vert })(2^{m} - 2^{\vert N\vert})$ & $2^{2m} - 2^{2m-\vert M\vert +1} + 2^{2m-2\vert M\vert }$\\
							\hline
							$(2^m - 2^{\vert M \vert -1})(2^m - 2^{\vert N \vert})$ & $2^{2m-\vert M\vert +1}-2^{2m-2\vert M\vert +1} - 2^{m+1} + 2^{m-\vert M \vert +1}$\\
							\hline
							$(2^{m-1} - 2^{\vert M \vert -1})(2^{m}- 2^{\vert N \vert})$ & $2^{m+1} - 2^{m-\vert M \vert +1}$\\
							\hline
							$0$ & $1$\\
							\hline
						\end{tabular}
					\end{adjustbox}
					\caption{Lee weight distribution in Theorem \ref{main} (\ref{proposition:4})}
					\label{table:4}
				\end{table}	
			 \item \label{proposition:5}
			 Let $D= a\Delta_{M} + c\Delta_N$ so that $D^{\textnormal{c}} =\big(a\Delta_{M}^{\textnormal{c}} + c\mathbb{F}_2^m\big)\bigsqcup \big( a\Delta_{M} + c\Delta_N^{\textnormal{c}}\big)$, where $\bigsqcup$ denotes disjoint union. Then $C^L_{D^{\textnormal{c}}}$ is a linear left-$E$-code of length $\vert D^{\textnormal{c}} \vert =2^{2m} - 2^{\vert M \vert + \vert N \vert}$ and size $2^{2m}$. The Lee weight distribution of $C^L_{D^{\textnormal{c}}}$ is displayed in Table \ref{table:5}.
			 \begin{table}[]
			 	\centering
			 	\begin{adjustbox}{width=\textwidth}
			 		\begin{tabular}{  c | c  }
			 			\hline
			 			Lee weight    & Frequency \\
			 			\hline
			 			$2^{2m}$ & $2^{2m-2\vert M\vert }-2^{m-\vert M\vert +1} +1$\\
			 			\hline
			 			$2^{2m-1}$ & $2^{m-\vert M\vert +1}-2$\\
			 			\hline
			 			$2^{2m}-2^{\vert M \vert + \vert N\vert}$ & $2^{2m} - 2^{2m-\vert M\vert +1} + 2^{2m-2\vert M\vert }$\\
			 			\hline
			 			$2^{2m} - 2^{\vert M \vert + \vert N \vert -1}$ & $2^{2m-\vert M\vert +1}-2^{2m-2\vert M\vert +1} - 2^{m+1} + 2^{m-\vert M \vert +1}$\\
			 			\hline
			 			$2^{2m-1} - 2^{\vert M \vert +\vert N \vert -1}$ & $2^{m+1} - 2^{m-\vert M \vert +1}$\\
			 			\hline
			 			$0$ & $1$\\
			 			\hline
			 		\end{tabular}
			 	\end{adjustbox}
			 	\caption{Lee weight distribution in Theorem \ref{main} (\ref{proposition:5})}
			 	\label{table:5}
			 \end{table}
			\end{enumerate}
		\end{theorem}
         \proof We discuss the proof of part (\ref{proposition:4}). The other parts can be proved in a similar way.\\         
         Let $x = a\alpha + c\beta \in E^m$. By Equation (\ref{keyeq1}) and (\ref{keyeq2}), we have
         \begin{equation}
         	\begin{split}
         		wt_{Lee}(c_{D}(x)) &= \vert D\vert -\frac{1}{2}(2^m-2^{\vert N\vert })\big[(2^m \delta_{0, \beta} - 2^{\vert M\vert }\Psi(\beta \vert M)) + (2^m \delta_{0, \alpha + \beta} - 2^{\vert M\vert }\Psi(\alpha + \beta \vert M))\big].
         	\end{split}
         \end{equation}
        Now we look at the following cases.
        \begin{enumerate}
        	\item If $\alpha =0, \beta =0 $, then $wt_{Lee}(c_D(x)) =0.$\\
            Here in this case, $\#\alpha =1, \#\beta =1$. Therefore, $\# x = 1$.
            \item If $\alpha \neq 0, \beta = 0 $, then
            \begin{equation*}
            	\begin{split}
                  wt_{Lee}(c_D(x)) &=\vert D\vert -\frac{1}{2}(2^m-2^{\vert N\vert })\big[(2^m  - 2^{\vert M\vert }) - 2^{\vert M\vert }\Psi(\alpha \vert M)\big].
            	\end{split}
            \end{equation*}
            \begin{itemize}
            	\item If $\Psi(\alpha \vert M) = 1 $ then $wt_{Lee}(c_D(x))   =(2^m-2^{\vert N \vert })\times 2^{m-1}.$\\
            	In this case, by using Lemma \ref{countingLemma},  we get\\
            	 $\#\alpha = (2^{m-\vert M \vert } -1)$,\\
            	$\#\beta = 1$.\\
            	 Therefore, $\#x = (2^{m-\vert M \vert } -1)$.
            	
            	\item If $\Psi(\alpha \vert M) = 0$ then $wt_{Lee}(c_D(x)) =\frac{1}{2}(2^{m} - 2^{\vert M \vert })(2^m-2^{\vert N \vert }).$\\ 
            	In this case, by using Lemma \ref{countingLemma},  we get \\
            	$\#\alpha =(2^{\vert M \vert }-1)\times 2^{m-\vert M\vert}$,\\
            	$\#\beta = 1$.\\
            	Therefore, $\#x= (2^{\vert M \vert }-1)\times 2^{m-\vert M\vert}$.            	
            	
            \end{itemize}
            
            \item If $\alpha = 0, \beta \neq 0 $, then
            \begin{equation*}
            	\begin{split}
            		wt_{Lee}(c_D(x)) &= (2^m-2^{\vert M \vert })(2^m-2^{\vert N \vert }) -\frac{1}{2}(2^m-2^{\vert N \vert})\big[-2^{\vert M\vert +1 }\Psi(\beta \vert M)\big].
            	\end{split}
            \end{equation*}
             \begin{itemize}
             	\item If $\Psi(\beta \vert M)= 1$ then $wt_{Lee}(c_D(x))=2^m\times (2^m - 2^{\vert N \vert }) $. In this case, by using Lemma \ref{countingLemma},  we get \\
             	$\#\alpha =1$,\\
             	$\#\beta = 2^{m-\vert M \vert} - 1$.\\
             	Therefore, $\#x= 2^{m- \vert M\vert } -1$.
             	
                \item If $\Psi(\beta \vert M)= 0$ then $wt_{Lee}(c_D(x))= (2^m - 2^{\vert M \vert})(2^m - 2^{\vert N \vert })$. In this case, by using Lemma \ref{countingLemma},  we get \\
             	$\#\alpha =1$,\\
             	$\#\beta = (2^{\vert M \vert } -1)\times 2^{m-\vert M \vert }$.\\
             	Therefore, $\#x= (2^{\vert M \vert } -1)\times 2^{m-\vert M \vert }$.
             \end{itemize}
        
            \item If $\alpha \neq 0, \beta \neq 0 $ and $\alpha = \beta $ ($\implies \alpha + \beta = 0$), then
            \begin{equation*}
            	\begin{split}
            		wt_{Lee}(c_D(x)) &=(2^m - 2^{\vert M \vert })(2^m - 2^{\vert N \vert }) -\frac{1}{2}(2^m-2^{\vert N\vert })\big[-2^{\vert M\vert }\Psi(\beta \vert M)+(2^m - 2^{\vert M \vert })\big].
            	\end{split}
            \end{equation*}
            \begin{itemize}
            	\item If $\Psi(\beta \vert M) = 1 $ then $wt_{Lee}(c_D(x)) = 2^{m-1}\times (2^m - 2^{\vert N \vert })$. In this case, by using Lemma \ref{countingLemma},  we get\\
            	$\#\alpha = 1$,\\
            	$\#\beta = 2^{m-\vert M\vert } -1$.\\
            	Therefore, $\#x=2^{m-\vert M\vert } -1$.
            	
            	\item If $\Psi(\beta \vert M) = 0$ then $wt_{Lee}(c_D(x))= \frac{1}{2}(2^m-2^{\vert M \vert })(2^m - 2^{\vert N \vert})$.
            	  In this case, by using Lemma \ref{countingLemma},  we get\\
            	  $\#\alpha = 1$,\\
            	  $\#\beta = (2^{\vert M\vert} -1)\times 2^{m-\vert M\vert }$.\\
            	  Therefore, $\#x = (2^{\vert M\vert} -1)\times 2^{m-\vert M\vert }$.
            	  
            \end{itemize}
    
        \item If $\alpha \neq 0, \beta \neq 0 $ and $\alpha \neq \beta $ ($\implies \alpha +\beta \neq 0$), then
        \begin{equation*}
        	\begin{split}
        		wt_{Lee}(c_D(x))&=(2^m-2^{\vert M\vert })(2^m-2^{\vert N \vert}) -\frac{1}{2}(2^m-2^{\vert N\vert })\big[-2^{\vert M\vert }\Psi(\beta \vert M) - 2^{\vert M\vert }\Psi(\alpha + \beta \vert M)\big].
        	\end{split}
         \end{equation*}
         \begin{itemize}
         	\item If $\Psi(\beta \vert M) = 0, \Psi(\alpha + \beta \vert M) = 0$ then $wt_{Lee}(c_D(x))= (2^m - 2^{\vert M \vert })(2^m - 2^{\vert N\vert })$. In this case, by using Lemma \ref{countingLemma},  we get\\
         	$\#\beta = (2^{\vert M \vert }-1)\times 2^{m-\vert M \vert}$,\\
         	$\#\alpha = (2^{\vert M\vert }-2)\times 2^{m-\vert M \vert } + (2^{m - \vert M\vert }-1)$.\\
         	Therefore, $\#x = \big((2^{\vert M\vert }-2)\times 2^{m-\vert M \vert } + (2^{m - \vert M\vert }-1)\big)(2^{\vert M \vert }-1)\times 2^{m-\vert M \vert}$
         	
         	\item If $\Psi(\beta \vert M) = 0, \Psi(\alpha + \beta \vert M) = 1$ then $wt_{Lee}(c_D(x))= (2^m - 2^{\vert M \vert -1 })(2^m - 2^{\vert N\vert }) $. In this case, by using Lemma \ref{countingLemma},  we get\\
         	$\#\alpha = (2^{\vert M \vert }-1)\times 2^{m-\vert M \vert} $,\\
         	$\#\beta = 1\times (2^{m-\vert M\vert }-1)$.\\
         	Therefore, $\#x= (2^{\vert M \vert }-1)\times 2^{m-\vert M \vert} \times (2^{m-\vert M\vert }-1)$.
         	
         	\item If $\Psi(\beta \vert M) = 1, \Psi(\alpha + \beta \vert M) = 0$ then $wt_{Lee}(c_D(x))= (2^m - 2^{\vert M \vert -1 })(2^m - 2^{\vert N\vert })$. In this case, by using Lemma \ref{countingLemma},  we get\\
         	$\#\alpha = (2^{\vert M \vert }-1)\times 2^{m-\vert M \vert}$,\\
         	$\#\beta = (2^{m- \vert M\vert }-1)$.\\
         	Therefore, $\#x= (2^{\vert M \vert }-1)\times 2^{m-\vert M \vert}\times(2^{m- \vert M\vert }-1)$.
         	
         	\item If $\Psi(\beta \vert M) = 1, \Psi(\alpha + \beta \vert M) = 1$ then $wt_{Lee}(c_D(x)) = 2^m\times (2^m-2^{\vert N \vert})$. In this case, by using Lemma \ref{countingLemma},  we get\\
         	$\#\alpha = (2^{m-\vert M \vert} -1)$,\\
         	$\#\beta = (2^{m-\vert M\vert } - 2)$.\\
         	Therefore, $\#x= (2^{m-\vert M \vert} -1)\times (2^{m-\vert M\vert } - 2)$.
         	
         \end{itemize}
         Based on the above calculations, we obtain Table \ref{table:4}.\\
         By using Table \ref{table:4}, $\vert \ker(c^L_D)\vert = \vert \{v\in E^m : v\cdot d = 0 ~\forall ~d \in D\}\vert =1$ so that $c^L_D$ is an isomorphism and hence $\vert C^L_{D}\vert = 2^{2m}$. \qed
        \end{enumerate}
      
        \subsection{Gray images of linear left-$E$-codes}
        In this subsection, we study the Gray images of linear left-$E$-codes in Theorem \ref{main}.\\
        Now we recall a result (see Theorem $1.4.8 ~(ii)$ of \cite{wchuffman}) for self-orthogonality of binary linear codes.
        \begin{theorem}\cite{wchuffman}\label{orthogonal_lemma}
        	If the Hamming weight of every non-zero element of a binary linear code $C$ is a multiple of $4$, then $C$ is self-orthogonal.
        \end{theorem}
        By using Theorem \ref{orthogonal_lemma}, we give a sufficient condition for the binary Gray images $\Phi(C^L_D)$ to be self-orthogonal for each $D$ discussed in Theorem \ref{main}.
        \begin{proposition}\label{sec4theoremLeft}
        	Let $m\in \mathbb{N}$ and let $M, N \subseteq [m]$. Assume that $D$ is as in Theorem \ref{main}. Then the parameters of the binary Gray image $\Phi(C^L_D)$ are given by Table \ref{table:minimalcondition} for each $D$. Besides, $\Phi(C^L_D)$ is a binary self-orthogonal code provided $\vert M \vert + \vert N \vert \geq 3$.
        	
        	\begin{table}[h]
        		\centering
        		\begin{adjustbox}{width=\textwidth}
        			\begin{tabular}{ c | c | c  }
        				\hline
        				S.N. &	$D$ as in&  $[n, k, d]$   \\
        				\hline
        				$1$	&	Theorem \ref{main}(\ref{proposition:1}) &  $[2^{\vert M \vert + \vert N \vert +1}, 2\vert M \vert, 2^{\vert M \vert + \vert N \vert -1}]$  \\
        				\hline
        				$2$	&	 Theorem \ref{main}(\ref{proposition:2}) &  $[(2^m - 2^{\vert M \vert }) 2^{\vert N \vert +1}, 2m, (2^m- 2^{\vert M \vert }) 2^{\vert N \vert -1}]$ \\
        				\hline
        				$3$	&	 Theorem \ref{main}(\ref{proposition:3}) &  $[2^{\vert M \vert +1} (2^m - 2^{\vert N \vert }), 2\vert M \vert, 2^{\vert M \vert -1}(2^m - 2^{\vert N \vert})]$ \\
        				\hline
        				$4$	&     Theorem \ref{main}(\ref{proposition:4}) & $[(2^{m+1} - 2^{\vert M \vert +1})(2^m - 2^{\vert N \vert}), 2m, (2^{m-1} - 2^{\vert M \vert -1})(2^m-2^{\vert N \vert})]$ \\
        				\hline
        				$5$ & Theorem \ref{main}(\ref{proposition:5})  & $[(2^{2m + 1}-2^{\vert M \vert + \vert N \vert +1}), 2m, (2^{2m - 1}-2^{\vert M \vert + \vert N \vert -1})]$\\
        				\hline
        			\end{tabular}
        		\end{adjustbox}
        		\caption{Parameters of $\Phi(C^L_D)$ for $C^L_D$ as in Theorem \ref{main}}
        		\label{table:minimalcondition}
        	\end{table}        	
        \end{proposition}
                
        The following examples illustrate Theorem \ref{main} and Proposition \ref{sec4theoremLeft}.
        
        \begin{example}
        	\begin{enumerate}
        		\item Let $m = 5$, $M = \{1, 2, 3\}, N=\{2, 3, 4\}$. Then $C^L_D$, where $D$ is as in Theorem \ref{main}(\ref{proposition:1}), is a linear left-$E$-code with Lee weight enumerator $X^{128} + 14 X^{96}Y^{32} + 49 X^{64} Y^{64}$. By Proposition \ref{sec4theoremLeft}, $\Phi(C^L_D)$ is a $[128, 6, 32]$ binary self-orthogonal code.
        		
        		\item Let $m = 4$, $M = \{2, 4\}, N=\{3\}$. Then $C^L_D$, where $D$ is as in Theorem \ref{main}(\ref{proposition:2}), is a linear left-$E$-code with Lee weight enumerator $X^{48} + 24 X^{36}Y^{12} + 72 X^{20} Y^{28} + 144 X^{24}Y^{24} + 6X^{32}Y^{16} + 9 X^{16}Y^{32}$. By Proposition \ref{sec4theoremLeft}, $\Phi(C^L_D)$ is a $[48, 8, 12]$ binary self-orthogonal code.
        		
        		\item Let $m = 4$, $M = \{1\}, N=\{2, 3, 4\}$. Then $C^L_D$, where $D$ is as in Theorem \ref{main}(\ref{proposition:3}), is a linear left-$E$-code with Lee weight enumerator $X^{32} + 2 X^{24}Y^{8} + X^{16} Y^{16}$. By Proposition \ref{sec4theoremLeft}, $\Phi(C^L_D)$ is a $[32, 2, 8]$ binary self-orthogonal code.
        		
        		\item Let $m = 5$, $M = \{1, 2\}, N=\{3, 4\}$. Then $C^L_D$, where $D$ is as in Theorem \ref{main}(\ref{proposition:4}), is a linear left-$E$-code with Lee weight enumerator $X^{1568} + 48 X^{1176}Y^{392} + 336 X^{728} Y^{840} + 576 X^{784}Y^{784} + 14 X^{1120}Y^{448} + 49 X^{672}Y^{896}$. By Proposition \ref{sec4theoremLeft}, $\Phi(C^L_D)$ is a $[1568, 10, 392]$ binary self-orthogonal code.
        		
        		\item Let $m = 4$, $M = \{1\}, N=\{3, 4\}$. Then $C^L_D$, where $D$ is as in Theorem \ref{main}(\ref{proposition:5}), is a linear left-$E$-code with Lee weight enumerator $X^{496} + 16 X^{372}Y^{124} + 112 X^{244} Y^{252} + 64 X^{248}Y^{248} + 14 X^{368}Y^{128} + 49 X^{240}Y^{256}$. By Proposition \ref{sec4theoremLeft}, $\Phi(C^L_D)$ is a $[496, 8, 124]$ binary self-orthogonal code.
        		
        	\end{enumerate}        	
        \end{example}
    
    \section{Linear right-$E$-codes using simplicial complexes}\label{section5}
    This section studies linear right-$E$-codes and their Gray images.\\	
    Let $m\in \mathbb{N}$ and let $D_i\subseteq \mathbb{F}_2^m, i=1, 2$. Assume that $D=aD_1+cD_2\subseteq E^m$. Define
    \begin{equation}\label{C_Ddefinitionright}
    	C^R_D=\{c^R_D(v)=\big(d\cdot v\big)_{d\in D} ~\vert~ v \in E^m\}
    \end{equation}
    where $x\cdot y=\sum\limits_{i=1}^{m}x_iy_i$ for $x,y\in E^m$.\\
    Then $C^R_D$ is a linear right-$E$-code of length $\vert D\vert $. The ordered set $D$ is called the \textit{defining set} of $C^R_{D}$.
    Observe that $c^R_{D}: E^m\longrightarrow C^R_{D}$ defined by $c^R_{D}(v)=\big(d\cdot v\big)_{d\in D}$ is an epimorphism of right-$E$-modules. 
    \subsection{Weight distributions of linear right-$E$-codes}
    
    Assume that $x=a\alpha + c\beta \in E^m$ and $d=at_1 +ct_2\in D$, where $\alpha, \beta \in \mathbb{F}_2^m$ and $t_i\in D_i, i=1, 2$. Then the Lee weight of $c^R_{D}(x)$ is 
    \begin{equation*}
    	\begin{split}
    		wt_{Lee}(c^R_{D}(x)) = 
    		  & wt_{L}\big(\big( \big(at_1 + ct_2 \big)\cdot \big(a\alpha +c \beta \big)\big)_{t_i\in D_i}\big)\\
    		= & wt_{Lee}\big(\big(a(\alpha t_1) + c(\alpha t_2)\big)_{t_i\in D_i}\big)\\
    		= & wt_{H}\big(\big(\alpha t_2\big)_{t_i\in D_i}\big) + wt_{H}\big(\big(\alpha t_1 + \alpha t_2\big)_{t_i\in D_i}\big)
    	\end{split}
    \end{equation*}   
    Now if $v \in \mathbb{F}_{2}^m$, then $wt_H(v)=0 \iff v=\textbf{0}\in \mathbb{F}_{2}^m$. Hence,
    \begin{equation}\label{keyeq1right}
    	\begin{split}
    		wt_{Lee}(c^R_{D}(x))
    		& =  \vert D \vert -\frac{1}{2}\sum\limits_{t_1\in D_1}\sum\limits_{t_2\in D_2}\big(1+(-1)^{\alpha t_2 } \big)\\
    		&+\vert D \vert -\frac{1}{2}\sum\limits_{t_1\in D_1}\sum\limits_{t_2\in D_2}\big(1+(-1)^{(\alpha t_1 +\alpha t_2)} \big)\\
    		&=\vert D\vert -\frac{1}{2}\sum\limits_{t_1\in D_1}(1)\sum\limits_{t_2\in D_2}(-1)^{\alpha t_2} - \frac{1}{2} \sum\limits_{t_1\in D_1}(-1)^{\alpha t_1 }\sum\limits_{t_2\in D_2}(-1)^{\alpha t_2}.
    	\end{split}
    \end{equation}
    Based on the above discussion, we have the following result.
     \begin{theorem}\label{mainright}
     	Suppose that $m\in \mathbb{N}$ and $M, N \subseteq [m]$.
     	\begin{enumerate}
     		\item \label{proposition:1right}
     		Let $D=a\Delta_M+ c\Delta_N\subseteq E^m$. Then $C^R_D$ is a $2$-weight linear right-$E$-code of length $\vert D\vert = 2^{\vert M\vert + \vert N \vert }$ and size $2^{\vert M\cup N \vert}$. The Lee weight distribution of $C^R_D$ is displayed in Table \ref{table:1right}.
     		\begin{table}[]
     			\centering
     			\begin{tabular}{  c | c  }
     				\hline
     				Lee weight    & Frequency \\
     				\hline
     				$2^{\vert M \vert + \vert N \vert }$ & $2^m\times (2^m-2^{m-\vert N\vert})$\\
     				\hline
     				$2^{ \vert M \vert + \vert N \vert -1}$ & $2^m \times (2^{m-\vert N \vert } - 2^{m-\vert M \cup N\vert })$\\     				
     				\hline
     				$0$ & $2^{2m-\vert M\cup N\vert}$\\
     				\hline
     			\end{tabular}
     			\caption{Lee weight distribution in Theorem \ref{mainright} (\ref{proposition:1right})}
     			\label{table:1right}	
     		\end{table}
        	\item \label{proposition:2right}
        	Let $D=a\Delta^{\textnormal{c}}_M+ c\Delta_N\subseteq E^m$. Then $C^R_D$ is a $3$-weight linear right-$E$-code of length $\vert D\vert = (2^m-2^{\vert M \vert})\times 2^{\vert N \vert }$ and size $2^{m}$. The Lee weight distribution of $C^R_D$ is displayed in Table \ref{table:2right}.
        	\begin{table}[]
        		\centering
        		\begin{tabular}{  c | c  }
        			\hline
        			Lee weight    & Frequency \\
        			\hline
        			$(2^m -2^{\vert M \vert })\times 2^{\vert N \vert }$ & $2^m \times (2^m - 2^{m-\vert N \vert})$\\
        			\hline
        			$2^{m + \vert N \vert -1}$ & $2^m \times (2^{m-\vert M \cup N \vert } - 1)$\\     				
        			\hline
        			$(2^m-2^{\vert M \vert })\times 2^{\vert N \vert -1}$& $2^m\times (2^{m - \vert N \vert } - 2^{m-\vert M \cup N \vert})$\\
        			\hline
        			$0$ & $2^{m}$\\
        			\hline
        		\end{tabular}
        		\caption{Lee weight distribution in Theorem \ref{mainright} (\ref{proposition:2right})}
        		\label{table:2right}	
        	\end{table}
            \item \label{proposition:3right}
            Let $D=a\Delta_M+ c\Delta^{\textnormal{c}}_N\subseteq E^m$. Then $C^R_D$ is a $3$-weight linear right-$E$-code of length $\vert D\vert = 2^{\vert M \vert}\times (2^m-2^{\vert N \vert})$ and size $2^{m}$. The Lee weight distribution of $C^R_D$ is displayed in Table \ref{table:3right}.
            \begin{table}[]
            	\centering
            	\begin{tabular}{  c | c  }
            		\hline
            		Lee weight    & Frequency \\
            		\hline
            		$2^{m+ \vert M \vert }$ & $2^m \times (2^{m-\vert M \cup N \vert}-1)$\\
            		\hline
            		$2^{\vert M \vert}\times (2^m - 2^{\vert N \vert -1})$ & $2^m \times (2^{m-\vert N \vert } - 2^{m-\vert M \cup N \vert})$\\     				
            		\hline
            		$2^{\vert M \vert }\times (2^m-2^{\vert N \vert })$ & $2^m\times (2^m - 2^{m - \vert N \vert })$\\
            		\hline
            		$0$ & $2^{m}$\\
            		\hline
            	\end{tabular}
            	\caption{Lee weight distribution in Theorem \ref{mainright} (\ref{proposition:3right})}
            	\label{table:3right}	
            \end{table}
            \item \label{proposition:4right}
            Let $D=a\Delta^{\textnormal{c}}_M+ c\Delta^{\textnormal{c}}_N\subseteq E^m$. Then $C^R_D$ is a $3$-weight linear right-$E$-code of length $\vert D\vert = (2^m-2^{\vert M \vert})(2^m -2^{\vert N \vert })$ and size $2^{m}$. The Lee weight distribution of $C^R_D$ is displayed in Table \ref{table:4right}.
            \begin{table}[]
            	\centering
            	\begin{tabular}{  c | c  }
            		\hline
            		Lee weight    & Frequency \\
            		\hline
            		$(2^m - 2^{\vert M \vert })(2^m -2^{\vert N \vert -1})$ & $2^m \times (2^{m-\vert N \vert}- 2^{m-\vert M \cup N \vert })$\\
            		\hline
            		$2^m(2^m - 2^{\vert M \vert }) - 2^{m + \vert N \vert -1}$ & $2^m \times (2^{m-\vert M\cup N \vert } - 1)$\\     				
            		\hline
            		$(2^{m} -2^{\vert M \vert })\times (2^m-2^{\vert N \vert })$ & $2^m\times (2^m - 2^{m - \vert N \vert })$\\
            		\hline
            		$0$ & $2^{m}$\\
            		\hline
            	\end{tabular}
            	\caption{Lee weight distribution in Theorem \ref{mainright} (\ref{proposition:4right})}
            	\label{table:4right}	
            \end{table}
            \item \label{proposition:5right}
            Let $D=a\Delta_M+ c\Delta^{\textnormal{c}}_N\subseteq E^m$ so that $D^{\textnormal{c}} = (a\Delta^{\textnormal{c}}_{M} + c\mathbb{F}_2^m)\bigsqcup (a\Delta_{M} + c\Delta^{\textnormal{c}}_{N})$ where $\bigsqcup$ denotes disjoint union. Then $C^R_D$ is a $3$-weight linear right-$E$-code of length $\vert D\vert = (2^{2m}-2^{\vert M \vert +\vert N \vert })$ and size $2^{m}$. The Lee weight distribution of $C^R_D$ is displayed in Table \ref{table:5right}.
            \begin{table}[]
            	\centering
            	\begin{tabular}{  c | c  }
            		\hline
            		Lee weight    & Frequency \\
            		\hline
            		$2^{2m}$ & $2^m \times (2^{m-\vert M\cup N \vert}- 1)$\\
            		\hline
            		$(2^{2m} - 2^{\vert M \vert + \vert N \vert -1})$ & $2^m \times (2^{m-\vert N \vert } - 2^{m-\vert M \cup N \vert})$\\     				
            		\hline
            		$(2^{2m} -2^{\vert M \vert + \vert N \vert})$ & $2^m\times (2^m - 2^{m - \vert N \vert })$\\
            		\hline
            		$0$ & $2^{m}$\\
            		\hline
            	\end{tabular}
            	\caption{Lee weight distribution in Theorem \ref{mainright} (\ref{proposition:5right})}
            	\label{table:5right}	
            \end{table}
        \end{enumerate}
     	
     \end{theorem}
        \subsection{Gray images of linear right-$E$-codes}
         Now, we study the Gray images of the codes in Theorem \ref{mainright}.\\
         By Theorem \ref{orthogonal_lemma}, we have several self-orthogonal binary codes.
         \begin{proposition}\label{mainright_Ortho}
         	Assume that $C^R_D$ is as in Theorem \ref{mainright}. Then its Gray image $\Phi(C^R_D)$ is a binary self-orthogonal code provided $\vert M \vert + \vert N \vert \geq 3$.
         \end{proposition}
         By using Lemma \ref{minimal_lemma} and Lemma \ref{GriesmerBound} respectively, we find minimal and optimal linear codes among the Gray images of codes in Theorem \ref{mainright}.
         \begin{theorem}\label{GrayImageC_DRight}
         	Suppose $\Phi$ is the map defined in Equation (\ref{PhiEq}). Let $m\in \mathbb{N}$ and $M, N\subseteq [m]$.
         	\begin{enumerate}
         	 \item \label{item_Right1}
         	 Let $D=a\Delta_{M} + c\Delta_{N}\subseteq E^m$. Then $\Phi(C^R_D)$ is a binary $[2^{\vert M \vert + \vert N \vert+1 }, \vert M \cup N \vert , 2^{\vert M \vert + \vert N \vert -1}]$-linear $2$-weight code.
         	 \item \label{item_Right2}
         	 Let $D=a\Delta^{\textnormal{c}}_{M} + c\Delta_{N}\subseteq E^m$. Then $\Phi(C^R_D)$ is a binary $[(2^m-2^{\vert M \vert })2^{\vert N \vert +1}, m, (2^m-2^{\vert M \vert})2^{\vert N \vert -1}]$-linear $3$-weight code.
         	 \item \label{item_Right3}
         	 Let $D=a\Delta_{M} + c\Delta^{\textnormal{c}}_{N}\subseteq E^m$. Then $\Phi(C^R_D)$ is a binary $[2^{\vert M \vert + 1 }(2^m - 2^{\vert N \vert}), m, 2^{\vert M \vert }(2^m - 2^{\vert N \vert})]$-linear $3$-weight code. If $\vert N \vert \leq m-2$ then $\Phi(C^R_D)$ is a minimal code.
         	 \begin{enumerate}
         	 	\item \label{item_Right3.1}
         	 	Let $\vert M \vert + \vert N \vert \leq m-1$ and $\theta_3 = 2^{\vert M \vert +1}-1$. If $1\leq \theta_3 < \vert M \vert + \vert N \vert +1$ then $\Phi(C^R_D)$ is optimal with respect to the Griesmer bound.
         	 	\item \label{item_Right3.2}
         	 	Let $m \leq \vert M \vert + \vert N \vert \leq 2m-1$ and $\theta_4 = 2^{\vert M \vert + \vert N \vert +1 -m}(2^{m-\vert N \vert} -1)$. If $0 < \theta_4 < m$ then $\Phi(C^R_D)$ is optimal with respect to the Griesmer bound.
         	 \end{enumerate}
         	 \item \label{item_Right4}
         	 Let $D=a\Delta^{\textnormal{c}}_{M} + c\Delta^{\textnormal{c}}_{N}\subseteq E^m$. Then $\Phi(C^R_D)$ is a binary $[2(2^m-2^{\vert M \vert})(2^m - 2^{\vert N \vert}), m, (2^m-2^{\vert M \vert})(2^m - 2^{\vert N \vert})]$-linear $3$-weight code. Moreover, it is a minimal code.
         	 \item \label{item_Right5}
         	 Let $D=a\Delta_{M} + c\Delta_{N}\subseteq E^m$ so that $D^{\textnormal{c}} = (a\Delta^{\textnormal{c}}_{M} + c\mathbb{F}_2^m)\bigsqcup (a\Delta_{M} +c \Delta^{\textnormal{c}}_{N})$. Then $\Phi(C^R_{D^{\textnormal{c}}})$ is a binary $[2(2^{2m} - 2^{\vert M \vert +\vert N \vert}), m, (2^{2m} - 2^{\vert M \vert +\vert N \vert})]$-linear $3$-weight code. If $\vert M \vert + \vert N \vert \leq 2m-2$ then $\Phi(C^R_{D^{\textnormal{c}}})$ is a minimal code.
         	\end{enumerate}
         \end{theorem}         
        The following examples illustrate Theorem \ref{mainright}, Proposition \ref{mainright_Ortho} and Theorem \ref{GrayImageC_DRight}.        
        \begin{example}
        	\begin{enumerate}
        		\item Let $m = 5$, $M = \{1, 2, 3\}, N=\{2, 3, 4\}$. If $D$ is as in Theorem \ref{mainright}(\ref{proposition:1right}), then $C^R_D$ is a linear right-$E$-code with Lee weight enumerator $X^{128} + X^{96}Y^{32} + 14X^{64} Y^{64}$. By Proposition \ref{mainright_Ortho} and Theorem \ref{GrayImageC_DRight}(\ref{item_Right1}), $\Phi(C^{R}_D)$ is a $[128, 4, 32]$ binary self-orthogonal code.
        		
        		\item Let $m = 5$, $M = \{1, 2\}, N=\{3, 4\}$. If $D$ is as in Theorem \ref{mainright}(\ref{proposition:2right}), then $C^R_D$ is a linear right-$E$-code with Lee weight enumerator $X^{224} + 6 X^{168}Y^{56} + X^{160} Y^{64} + 24 X^{112} Y^{112}$. By Proposition \ref{mainright_Ortho} and Theorem \ref{GrayImageC_DRight}(\ref{item_Right2}), $\Phi(C^{R}_D)$ is a $[224, 5, 56]$ binary self-orthogonal code.
        		
        		\item 
        		\begin{enumerate}
        			\item Let $m = 5$, $M = \emptyset, N=\{1, 2, 3\}$. If $D$ is as in Theorem \ref{mainright}(\ref{proposition:3right}), then $C^R_D$ is a linear right-$E$-code with Lee weight enumerator $X^{48} + 28 X^{24}Y^{24} + 3 X^{16} Y^{32}$. By Proposition \ref{mainright_Ortho} and Theorem \ref{GrayImageC_DRight}(\ref{item_Right3.1}), $\Phi(C^{R}_D)$ is a $[48, 5, 24]$ binary minimal and self-orthogonal code. Note that Gray image $\Phi(C^{R}_D)$ is optimal according to the Database \cite{BoundTable}.
        			
        			\item Let $m = 9$, $M = \{1, 2\}, N=\{2, 3, 4, 5, 6, 7, 8\}$. If $D$ is as in Theorem \ref{mainright}(\ref{proposition:3right}), then $C^R_D$ is a linear right-$E$-code with Lee weight enumerator $X^{3072} + 508 X^{1536}Y^{1536} + 2 X^{1280} Y^{1792} + X^{1024}Y^{2048}$. By Proposition \ref{mainright_Ortho} and Theorem \ref{GrayImageC_DRight}(\ref{item_Right3.2}), $\Phi(C^{R}_D)$ is a $[3072, 9, 1536]$ binary minimal and self-orthogonal code. Note that Gray image $\Phi(C^{R}_D)$ is optimal as there doesn't exist $[3072, 9, 1537]$-linear code over $\mathbb{F}_2$.
        		\end{enumerate}
        		
        		\item Let $m = 4$, $M = \{1\}, N=\{2, 3\}$. If $D$ is as in Theorem \ref{mainright}(\ref{proposition:4right}), then $C^R_D$ is a linear right-$E$-code with Lee weight enumerator $X^{336} + 12 X^{168}Y^{168} + X^{144} Y^{192} + 2 X^{140}Y^{196}$. By Proposition \ref{mainright_Ortho} and Theorem \ref{GrayImageC_DRight}(\ref{item_Right4}), $\Phi(C^{R}_D)$ is a $[336, 4, 168]$ binary minimal and self-orthogonal code.
        		
        		\item Let $m = 4$, $M = \{1, 2, 3\}, N=\{2, 3, 4\}$. If $D$ is as in Theorem \ref{mainright}(\ref{proposition:5right}), then $C^R_D$ is a linear right-$E$-code with Lee weight enumerator $X^{384} + 14 X^{192}Y^{192} + X^{160} Y^{224}$. By Proposition \ref{mainright_Ortho} and Theorem \ref{GrayImageC_DRight}(\ref{item_Right4}), $\Phi(C^{R}_D)$ is a $[384, 4, 192]$ binary minimal and self-orthogonal code.
        	\end{enumerate}
        \end{example}
    \begin{remark}
    	By Theorem \ref{Bonisoli} all the binary $1$-weight codes are simplex codes and hence they are distance optimal.
    \end{remark}
    \begin{table}[h]
    	\centering
    	\begin{adjustbox}{width=\textwidth}
    		
    		\begin{tabular}{c|c|c|c|c}
    			\hline
    			 	Result & $[n,k,d]$-code & \#Weight & Distance optimal & Minimal  \\
    			\hline    			 
    			Theorem \ref{GrayImageC_DRight}(\ref{item_Right3.1}) & \multirow{2}{*}{$[2^{\vert M \vert + 1 }(2^m - 2^{\vert N \vert}), m, 2^{\vert M \vert }(2^m - 2^{\vert N \vert})]$} & \multirow{2}{*}{3} & Yes, if $1\leq \theta_3 < \vert M \vert + \vert N \vert +1 \leq m$ & \multirow{2}{*}{Yes, if $\vert N \vert \leq m-2$} \\
    			\cline{1-1}\cline{4-4}
    			Theorem \ref{GrayImageC_DRight}(\ref{item_Right3.2})  &  &  & Yes, $0 < \theta_4 < m \leq \vert M \vert + \vert N \vert$ & \\ \hline
    			Theorem \ref{GrayImageC_DRight}(\ref{item_Right4}) & $[2(2^m-2^{\vert M \vert})(2^m - 2^{\vert N \vert}), m, (2^m-2^{\vert M \vert})(2^m - 2^{\vert N \vert})]$ & 3 & $-$ & Yes \\
    			\hline
    			Theorem \ref{GrayImageC_DRight}(\ref{item_Right5}) 	& $[2(2^{2m} - 2^{\vert M \vert +\vert N \vert}), m, (2^{2m} - 2^{\vert M \vert +\vert N \vert})]$ & 3 & $-$ & Yes, if $\vert M \vert + \vert N \vert \leq 2m-2$ \\
    			\hline
    		\end{tabular}
    		
    	\end{adjustbox}
    	\caption{Binary linear codes from Simplicial complexes in this article}
    	\label{table:article}
    	
    \end{table}
		\section{Conclusion}\label{section6}
		Linear left-$E$-codes and right-$E$-codes denoted, respectively, by $C_D^L$ and $C_D^R$ are studied using simplicial complexes having one maximal element. This is the first attempt to obtain linear codes over a non-unital non-commutative ring using such a construction. We use a Gray map to study the corresponding binary codes for both $C_D^L$ and $C_D^R$. We obtain their weight distributions; Boolean functions are used in these computations. As a consequence, we achieve two infinite families of optimal codes. Besides, we obtain many families of binary minimal codes. Most of the codes obtained in this article are few-weight codes. All the binary codes obtained here are self-orthogonal under certain mild conditions. For the reader's convenience, we list certain few-weight codes having good parameters in Table \ref{table:article} obtained in this article.
						

	\end{document}